\documentclass[10pt,conference]{IEEEtran}
\IEEEoverridecommandlockouts

\usepackage{tabularx, makecell, multirow, float, stfloats}
\usepackage{xcolor, colortbl}
\usepackage[square, sort, numbers]{natbib}
\usepackage{graphicx}
\usepackage{balance, xurl}
\usepackage[redeflists]{IEEEtrantools}
\usepackage{tcolorbox}
    
\definecolor{lightgray}{rgb}{0.8,0.8,0.8}

\newcolumntype{s}{>{\hsize=.25\hsize}X}

\begin{document}
\bstctlcite{IEEEexample:BSTcontrol}

\title{Technical Debt in the Peer-Review Documentation of R Packages: a rOpenSci Case Study
\thanks{This work is partially supported by Natural Sciences and Engineering Research Council of Canada RGPIN-2019-05175.}}

\author{
\IEEEauthorblockN{Zadia Codabux}
\IEEEauthorblockA{\textit{University of Saskatchewan}}
\IEEEauthorblockA{\textit{zadiacodabux@ieee.org}}
\and
\IEEEauthorblockN{Melina Vidoni}
\IEEEauthorblockA{\textit{RMIT University}}
\IEEEauthorblockA{\textit{melina.vidoni@rmit.edu.au}}
\and
\IEEEauthorblockN{Fatemeh H. Fard}
\IEEEauthorblockA{\textit{University of British Columbia}}
\IEEEauthorblockA{\textit{fatemeh.fard@ubc.ca}}
}

\maketitle
\thispagestyle{plain}
\pagestyle{plain}

\begin{abstract}
\textit{Context:} Technical Debt (TD) is a metaphor used to describe code that is ``not quite right." Although TD studies have gained momentum, TD has yet to be studied as thoroughly in non-Object-Oriented (OO) or scientific software such as R. R is a multi-paradigm programming language, whose popularity in data science and statistical applications has amplified in recent years. Due to R's inherent ability to expand through user-contributed packages, several community-led organizations were created to organize and peer-review packages in a concerted effort to increase their quality. Nonetheless, it is well-known that most R users do not have a technical programming background, being from multiple disciplines. \textit{Objective:} The goal of this study is to investigate TD in the documentation of the peer-review of R packages led by \textit{rOpenSci}. \textit{Method:} We collected over 5,000 comments from 157 packages that had been reviewed and approved to be published at rOpenSci. We manually analyzed a sample dataset of these comments posted by package authors, editors of \textit{rOpenSci}, and reviewers during the review process to investigate the types of TD present in these reviews. \textit{Results:} The findings of our study include (i) a taxonomy of TD  derived from our analysis of the peer-reviews (ii) documentation debt as being the most prevalent type of debt (iii) different user roles are concerned with different types of TD. For instance, reviewers tend to report some types of TD more than other roles, and the types of TD they report are different from those reported by the authors of a package. \textit{Conclusion:} TD analysis in scientific software or peer-review is almost non-existent. Our study is a pioneer but within the context of R packages. However, our findings can serve as a starting point for replication studies, given our public datasets, to perform similar analyses in other scientific software or to investigate the rationale behind our findings.\\ 
\end{abstract}


\section{Introduction}
Technical Debt (TD) describes the problem of balancing near-term value with long-term quality \cite{Ernst2015}. The relevance of this metaphor in software development has been widely acknowledged, and researchers have studied this phenomenon from different perspectives \cite{Bavota2016}. Nonetheless, the artifacts used to study TD have remained somewhat the same \cite{SIERRA201970}, with only recent efforts venturing beyond source code \cite{Bellomo2016, Laerte2020}. Managing technical debt is crucial as software containing debt are often more prone to defects, changes, vulnerabilities, and other software issues \cite{Khomh, hall2014some, codabux2017, codabux2017relationship, sultana2020}.

The role of software in data science and statistics has increased considerably over the years \cite{Howison2011}. As a result, scientific organizations identified the production and maintenance of scientific software as an area of concern, as it can affect the validity of results \cite{stewart2010editors}. The relevance of this issue has increased recently, as top conferences have started reviewing the reproducibility and transparency of software tools used in academic production\footnote{\url{https://bit.ly/2IsRFoW}}. However, the process of reviewing scientific production is demanding, requiring extensive amounts of human effort \cite{Balachandran2013}. Moreover, regardless of its popularity, there is not much research in R from a software engineering perspective, and technical debt has not been explored in this context \cite{Ihaka2017}. Additionally, most R programmers do not have a solid programming background like a Software Engineer may have \cite{C10}. 

One of the most widely used programming languages for scientific software is R. It is a multi-paradigm language that has grown considerably due to its package-based structure and developers' ability to contribute packages publicly available as extensions of the ``base" language \cite{Theubl2011}. It is commonly used in scientific applications, such as statistics and data science \cite{Decan2016}.  CRAN, the Comprehensive Archive Network, was created for users to suggest improvements and report bugs and, nowadays, it is the official venue to submit user-generated R packages \cite{Ihaka2017}. However, although it has a growing community, most package contributors are not software engineers by training \cite{C10}, and only a few of them are mindful of the inner concepts of the language \cite{excel15}. This lack of formal programming training can lead to lower code quality \cite{Gilson2020}.

Several community-led organisations were created to organize and review packages - among them, \textit{rOpenSci} \cite{ropensci2015} and \textit{BioConductor} \cite{Gentleman2004}. In particular, \textit{rOpenSci} has established a thorough peer-review process for R packages, based on the intersection of academic peer-reviews and software reviews. \textit{rOpenSci} is a well-known organization that conducts a thorough peer-review process, which has become an integral part of the R community \cite{ropensci2015}. Moreover, they have forged partnerships with academic periodicals to complete their review process\footnote{\url{https://ropensci.org/blog/2017/11/29/review-collaboration-mee/}}. Their peer-review process is also open (i.e. not blinded\footnote{\url{https://devguide.ropensci.org/}}), and conducted in public GitHub issues\footnote{\label{fn:issues}\url{https://github.com/ropensci/software-review/}}. \textit{rOpenSci}'s scope is far more extensive than \textit{BioConductor}'s. The latter is limited to specific bioinformatics packages only and interacts with existing packages and reuses their data structures\footnote{\url{https://www.bioconductor.org/developers/package-submission/}}. Moreover, although CRAN allows for the submission of packages, it does not perform a thorough review of the submissions. Instead, it opts for generalized and partially automated checks.

This study aims to investigate TD  in scientific software reviews, specifically in the context of R packages. We focused on the \textit{rOpenSci}'s open peer-review process, which is handled using GitHub issues. We extracted 157 completed and accepted package reviews, consisting of more than 5,000 individual comments. The 157 reviews were broken down into individual comments and classified by the comment's author's role in the peer-review process. A sample of 358 comments, further broken down into 600 phrases, was inspected and manually analyzed by two authors to extract TD instances. 

Our findings indicate: 
\begin{itemize}
    \item Several types of TD were found in the peer-review documentation, and the definitions were adjusted to fit the inherent characteristics of the R programming language. Furthermore, the types of debt were grouped according to three perspectives (namely users, developers and CRAN), representing ``who" is the most affected by the debt type.
    
    \item Regardless of the participants' roles, \textit{documentation} is the most prominent and \textit{versioning} is the least encountered type of debt in the peer-review process. 
    
    \item The participants (or ``user roles" as referred to in this study) of the review process (i.e. \textit{authors}, \textit{editors} of \textit{rOpenSci} and \textit{reviewers} of the submitted packages) appear to focus on different types of debt. 
    The percentage of the phrases with \textit{documentation debt} is the highest among the \textit{reviewers} and \textit{editors}. However, \textit{authors} tend to report more \textit{defect debt}. Moreover, the distribution of debt types found in the comments is significantly different between \textit{reviewers}, and other roles, while the distribution of TD types are not different among the comments written by \textit{authors} or \textit{editors}. Furthermore, the \textit{reviewers} reported most of the TD instances. 
\end{itemize}

Overall, the contributions of this study are:
\begin{itemize}
    \item This is the first study that investigates TD in the R package peer-reviewing documentation. We use a new source of data to explore TD, compared to most studies from the literature which analyze source code.  
    \item We propose a taxonomy of TD, extended to incorporate specific concepts related to R packages.
    \item A dataset of manually labelled TD instances in R  peer-reviews documentation is made available publicly. 
\end{itemize}

\textit{Paper Structure}. Section \ref{sec:relatedWork} highlights related works on TD in traditional software and data science applications, and R-related research. Next, Section \ref{sec:methodology} describes our methodology by outlining our goal and research questions, discussing the dataset's construction, and elaborating on the data analysis. Section \ref{sec:results} presents the results of our research questions, and Section \ref{sec:implications} discusses the implications of our results for the broader area of R programming and TD. Finally, Section \ref{sec:threats} acknowledges the threats to the validity of the study, and Section \ref{sec:conclusions} concludes this work.

\section{Related Work \label{sec:relatedWork}}
This section gives an overview of the relevant literature regarding TD  taxonomy and classification, TD  in data science, and software engineering research in R.

\textbf{Technical Debt Taxonomy.}
The most common types of TD  include \textit{code, design, architecture, documentation, test (and test automation),} and \textit{defect} \cite{li2015systematic}, referring to debt in the various phases of the software development life cycle. However, despite being criticised for ``diluting" the TD  metaphor \cite{kruchten2012technical}, other aspects of software that causes issues in the software development process have created additional types of debt, namely, \textit{build} \cite{morgenthaler2012searching}, \textit{service} \cite{alzaghoul2013cloudmtd}, \textit{versioning} \cite{greening2013release}, \textit{usability} \cite{zazworka2013case}, \textit{people} \cite{kruchten2012technical}, \textit{process} \cite{codabux2017empirical}, \textit{social} \cite{tamburri2013social}, \textit{database (design)} \cite{al2016database}, \textit{environmental} \cite{magazine2016technical}, \textit{data} \cite{codabux2017empirical}, and \textit{infrastructure} \cite{debois2008agile}. However, according to surveys of the literature on TD, these debts have generated interest among the research community and have been adopted as additional types of TD  in the software development process \cite{li2015systematic, alves2014towards}.

\textbf{Technical Debt in Data Science Applications.} TD  in Machine Learning (ML) systems was first discussed by Sculley et al. \cite{sculley2015hiddenTDinML}. They explored multiple hidden TD in ML systems, including entanglement, data dependencies and configuration issues, which can adversely impact the system design. Breck et al. \cite{breck2017ml} proposed specific tests to reduce TD and ensure ML systems' production readiness. Self-Admitted Technical Debt (SATD) in deep learning frameworks has been investigated to show that the frameworks contain \textit{design debt} as the major TD type, followed by debts in the \textit{defect, documentation, test, requirement, compatibility,} and \textit{algorithm} categories \cite{liu2020usingSATDinDL}. 
Although not explicitly mentioning TD, several studies have developed guidelines for the quality assurance of machine learning systems \cite{hamada2020guidelines}, developed tools for data validation \cite{polyzotis2019data}, or studied the software engineering practices for machine learning \cite{amershi2019software, wan2019does}. 

\textbf{Software Engineering Research in R Programming.} A recurrent topic of research in R is the analysis of package and dependency networks, e.g., dependency management and the impact of GitHub in non-CRAN packages \cite{Decan2016}, measuring package activity and lifecycle \cite{PLAKIDAS2017119} and the growth and maintainability capabilities of CRAN packages \cite{Claes2014}. Another popular area, given R's mix of programming paradigms, is programming language theory, addressing problems such as code profiling \cite{Rubio2015}, wrappers to other languages (e.g., C++) \cite{JSSv040i08}, type systems \cite{Turcotte2019}, and lazy evaluation \cite{Fluckinger2019}. 

Several studies have been conducted in traditional software code reviews \cite{Balachandran2013, Zanaty2018} but only a few have focused on scientific software reviews \cite{Howison2011, Kanewala2019, Gruning2019}. However, this study is focused on R's peer review documentation and code review is outside the scope of this work. 

The study of TD in machine learning goes back to 2015 but has since gained popularity, with recent studies on SATD in deep learning applications. Although R is one of the main languages used in data science applications, we are not aware of any research investigating TD in the R language. Besides, many of the TD-related researches explore the TD in source code. However, in this study, we are exploring TD in the documentation of R packages' peer-reviews. 

\section{Methodology \label{sec:methodology}}
\subsection{Goal and Research Questions}
This study investigates the presence of TD  in the documentation of the peer-review process of scientific software, particularly in the \textit{rOpenSci} case. Specifically, this led us to the following research questions:

\begin{itemize}
\item \textbf{RQ1: What types of TD are analyzed during the peer-review process?} The rationale behind this question is to determine the different types of TD that will emerge from the peer-review process documentation, as reported by the different user roles. We will use the types of TD  associated with traditional software (as listed in Section \ref{sec:relatedWork}) as a starting point to build a taxonomy specific to scientific software. 

\item \textbf{RQ2: What is the distribution of TD according to the types identified in RQ1?} We want to determine the distribution of the different types of TD in the documentation of the peer-reviews and analyze which debt is the most and least prevalent in the peer-reviews. 

\item \textbf{RQ3: Are identified TD types independent of the user roles? } 
We investigate the association between the user roles (as described in Table \ref{tab:author_roles}) and TD types. We explore the differences among the TD types identified for each user role. This will give us insights into the differences among user roles in terms of the TD types they report. 
\end{itemize}

\subsection{Data Extraction}

\begin{figure*}[t!]
    \centering
    \includegraphics[width=\textwidth]{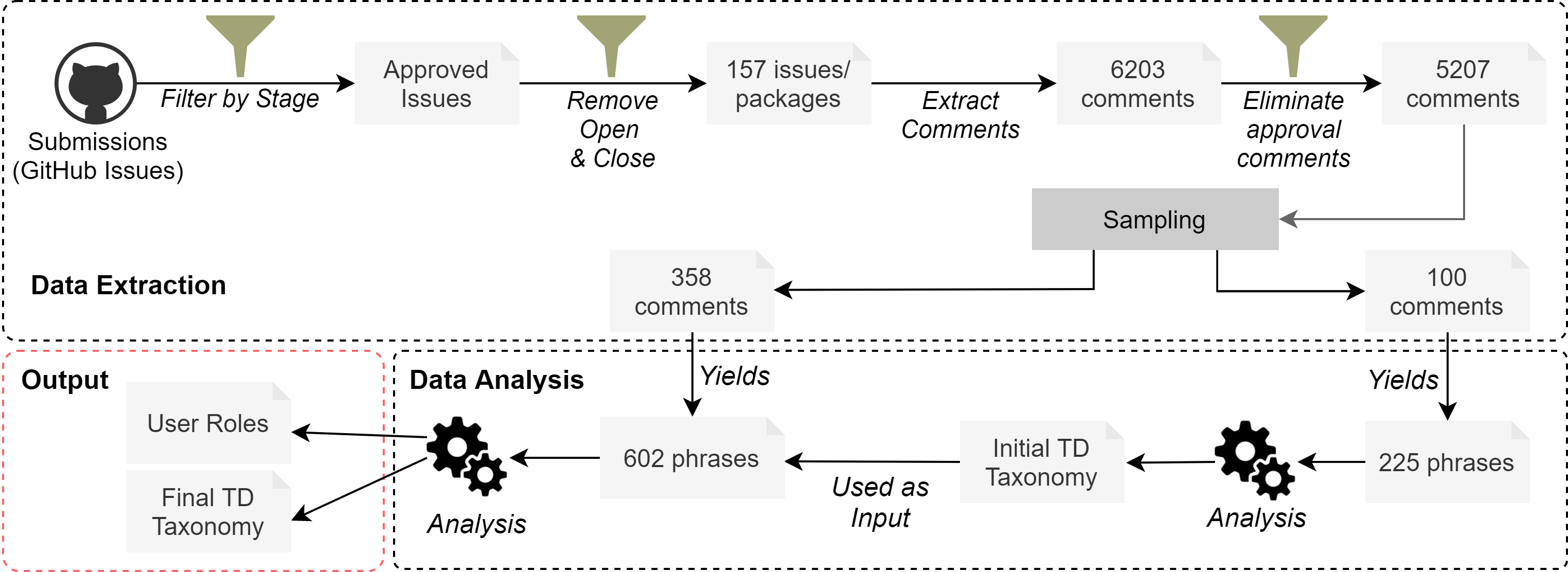}
    \caption{ Schematic Diagram of the Study}
    \label{fig:block_diag}
\end{figure*}

\textit{rOpenSci} handles every submission to be peer-reviewed as a GitHub issue in their Software Review Repository\footnotemark\protect[\ref{fn:issues}]. For more information about \textit{rOpenSci}'s review process, the Peer-Review Guide is openly available\footnote{\url{https://devguide.ropensci.org/softwarereviewintro.html}}. Overall, there are two types of submissions: \textit{pre-submissions} (in which the authors enquire about the fit of the proposed package to \textit{rOpenSci}'s scope), and \textit{submissions} (referring to the software peer-review process) \cite{ropensci2015}. For this study, the latter was considered. GitHub ``labels'' are used to indicate the stage in which the review is at.

A submission starts with an \textit{author} creating a new GitHub issue using the ``package submission'' template\footnote{\url{https://bit.ly/38sR7bt}}; after some initial checks and questions from the \textit{handling editor}, at least two \textit{reviewers} are selected and invited by the \textit{handling editor}. The \textit{reviewers} must have experience both on the topic and R. Each of them needs to conduct a review of the package and submit their revision using the review template\footnote{\label{fn:reviewTemplate}\url{https://devguide.ropensci.org/reviewtemplate.html#reviewtemplate}}. The process continues as a back-and-forth communication between \textit{reviewers} and \textit{author}, moderated by the \textit{handling editor}. Once the package is approved, the handling editor posts an approval comment\footnote{\label{fn:approvalTemplate}\url{https://devguide.ropensci.org/approvaltemplate.html}} and the issue is closed afterwards. Additional communication related to the administrative on-boarding process, but not any package discussion, may happen. 

A label identifies whether a submitted package for peer-review is still \texttt{under review} (and in which step of the process), \texttt{withdrawn}, \texttt{on-hold} or \texttt{approved} to be published by \textit{rOpenSci}. The approved packages are the ones for which the review process is completed, and the package authors have applied the feedback they have received. 
We developed an R script using the GitHub API and mined the reviews from the GitHub issues of the approved submissions. This process is described next and can be visualized in the ``Data Extraction" phase of Figure \ref{fig:block_diag}.

First, we extracted issues tagged as \texttt{approved}. This indicates that an issue is a \textit{submission} for peer-review and its review has been completed and the package is approved for ``onboarding'' (i.e. akin to ``publication'') in \textit{rOpenSci}. The packages with \texttt{open} or \texttt{close} tags were disregarded, as \textit{rOpenSci} issues can remain open for reasons unrelated to the review. Following this, we extracted \textbf{157 peer-review issues to be studied}, which correspond to 157 different packages. This dataset was converted to a CSV file, which contains data about: \emph{labels assigned, the issue opening comments, dates for creation, update and closing, number of comments of the issue, the handle of the opening author, title, issue number, issue ID and URL}. Every row of this dataset belongs to a single issue.

The 157 approved issues have a timestamp for the creation and the time the issue is closed. We use the difference between these two timestamps to determine how many days it takes for packages to be reviewed and approved after they are submitted. The median duration for issue approvals is 104.7 days, with the 0.25 percentile of 69.7 days and 0.75 percentile of 193.1 days. In terms of outliers, there are five packages for which the duration is between 400 and 550 days. 

Second, we used the unique \textit{issue number} in combination with the GitHub API to extract the comments for each issue. It should be noted that there are multiple comments associated with each issue, posted by package authors, reviewers, or editors of \textit{rOpenSci}. The minimum number of comments per issue is one, the maximum is 113, and the median is 37. We obtained \textbf{6203 individual comments for this complementary dataset}. In particular, it includes the following fields: \emph{unique comment ID and URL, issue number, dates for create and update, user handle of the author, and the comment body}. Here, each row belongs to a comment, and multiple rows may link back to the same GitHub issue.

It is worth mentioning that: 
\begin{itemize}
\item Selected reviews were not discriminated by year, considering issues from \textit{rOpenSci}'s peer-review dates back to 2011; however, the early reviews (about the first six months since \textit{rOpenSci}'s foundation) did not have a well-defined peer-review process. 

\item \textit{rOpenSci} uses a pre-built template for issue openings, where the authors have to input specific data, akin to the submission form in an academic journal. Therefore, it was established that these opening posts did not contain any discussion regarding the package and were excluded from the analysis.
\end{itemize}

Often, the discussion may continue after the package's acceptance; however, these comments are related to the ``publication'' of the package in \textit{rOpenSci}, as well as its promotion on different academic networks. As a result, these discussions do not focus on the quality of the code and do not include any additional TD instances.

The comments were filtered in a two-fold process: first, an automated search for the ``approval template,''\footnotemark\protect[\ref{fn:approvalTemplate}] eliminating all ``default'' and second, a manual search eliminating comments indicating approval without using the template. Although the ``approval template'' is suggested by \textit{rOpenSci}, it is not enforced, and editors may customize it. This process reduced the comments to 5207.

Lastly, we sampled\footnote{\url{https://www.surveysystem.com/sscalc.htm}} a subset of the 5207 comments with a confidence level of 95\% and confidence interval of 5, resulting in 358 comments.  This sample is used for manual investigation and further analysis. We extracted a random sample of 100 comments initially as a learning phase for RQ1 before analyzing the 358 comments. The two authors who applied the manual investigation have previous experience with TD. One of the authors has extensive research experience with TD (8+ years), and the other is an R expert with research experience in TD.

\subsection{Data Analysis\label{sec:agreement}}
In this section, we describe how we extracted instances of TD  and user roles from the comments.  This process can be visualized in the ``Data Analysis" phase of Figure \ref{fig:block_diag}. 

\subsubsection{Comment Classification}
To classify the phrases, we used a \emph{card sorting technique} \cite{whitworth2006encyclopedia}. Card sorting is commonly used to derive taxonomies from data. Our card sort was closed, that is, we used a pre-defined list of TD  types. We extracted the different types of TD  from the literature and compiled a list of the 17 types of debt, as enumerated in Section \ref{sec:relatedWork}. 

We performed two iterations to classify the comments. 
In the first iteration, we randomly picked 100 comments from the 5207 comments. This phase resulted in an initial taxonomy of 10 different types of TD.
The purpose of this phase is two-fold: (1) familiarizing the authors with the R terminologies and the peer-review process, (2) extracting an initial taxonomy of TD. 
In the first iteration, out of the 100 comments, 36 had TD and were further broken down into 225 phrases for analysis. 
The rationale for breaking the comments into phrases is that the comments were rather lengthy and contained different TD types. Each comment was broken down first into paragraphs and then into sentences (by either checking for a period or a semi-colon). If two sentences (or more) were related and referred to the same point, they were kept together as one phrase.
Two of the authors separately manually classified them according to the 17 types of TD  using the closed card sorting technique. Out of the 225 phrases, the manual classification led to 44 disagreements. They discussed the disagreements and finalized the classification of these 44 phrases. This first iteration reduced the list of 17 types of TD to 10, namely: \textit{architecture, build, code, defect, design, documentation, requirements, test, usability and versioning}. 

In the second iteration, we randomly sampled 358 comments from the remaining in the documentation, ensuring that they do not have an intersection with the sample of 100 used in the first phase. Out of the 358 sampled comments, we manually extracted 95 that contained TD and discarded the remaining 264 with no TD. The 95 comments were broken down into 602 phrases. The first two authors manually classified the phrases, but this time, according to the taxonomy in the first iteration. If a phrase could not be classified, it is marked with a question mark to be discussed later. Then, the authors discussed all of the disagreements and the cases they were unsure about its category. In the discussions, the authors decided on the final classification of the disagreements. During the second iteration, no additional types of debt were uncovered, suggesting that the 100 comments were pretty representative of the dataset regarding the different types of TD present. Therefore, the taxonomy of TD  was finalized. Two phrases out of the 602 were removed as the authors could not decide whether it was a debt or not, yielding a total of 600 phrases that were classified. 

Revision ``comments" in \textit{rOpenSci} are large since the process is derived from academic peer-review; they are organized in paragraphs and sentences. Each comment was broken down into paragraphs and then into sentences (by either checking for a period or a semi-colon). If two sentences (or more) were related and referred to the same point, they were kept together as a \textit{phrase}. Here, the word `comment' refers to GitHub's `comment' on an issue (where the review process is conducted) and is not representative of the text's length.

\begin{table}[h!]
  \caption{Assigned roles and the process used to extract them}
  \label{tab:author_roles}
  \begin{tabularx}{0.5\textwidth}{XX}
\hline
Role  & Source \\
\hline
\textbf{op\_author.} Equivalent to a corresponding author. & Assigned to the author that opened the issue. This is extracted from the dataset during the Data Extraction phase. \\

\textbf{handling\_editor}. The associate editor handling the issue in the same sense as in academic journals. & \textit{rOpenSci} assigns (in GitHub terms) this user to the issue. This was extracted from the first dataset and automatically labelled. \\

\textbf{author}. Other authors or maintainers of the package. & Often mentioned in the opening post, they were only labelled if their GitHub handles were mentioned. The assigned roles of the GitHub handles were extracted by reading all the comments for an issue. \\

\textbf{editors}. Other editors, such as the Editor in Chief & They were manually extracted from the dataset; this role often checks the scope suitability of the package and assigns a \textit{handling\_editor}.\\

\textbf{reviewers}. The experts assigned as referees for the submission. & They were found by inspecting the \textit{handling editor} post, in which these are often mentioned, and by inspecting the reviews in search for the ``review template''\footnotemark\protect[\ref{fn:reviewTemplate}].\\
\hline    
    \end{tabularx}
\end{table}

Then, we calculated the inter-rater reliability using Cohen's Kappa coefficient \cite{mchugh2012interrater}. 
Cohen's Kappa is a test that measures the raters' agreement in studies that have two or more raters responsible for measuring a categorical scale variable. The test results in a number between $-1$ and $+1$, where $-1$ shows the highest disagreement and $+1$ shows the most agreement rate. The threshold cutoff for deciding on the high agreement varies based on the fields  \cite{mchugh2012interrater}. We use the score of above $0.79$ as a high agreement rate, which is used in previous and similar studies in software engineering \cite{liu2020usingSATDinDL, huang2018identifying}. We obtain the result of $+0.80$, which indicates the high rate of agreement and reliability of our coding schema.   

\subsubsection{User Roles}
Each comment in the collected dataset is written by a user (identified by GitHub handles), and each user has a different role. The values of the user roles are either pre-defined or null. In this step, for each comment of an issue, we extract the user roles for each of the user handles, as the same handles can have different roles in multiple issues. For example, a user can be an \textit{author} of a package in one issue and a \textit{reviewer} of another package.  The pre-defined roles and their extraction process, are summarised in Table \ref{tab:author_roles}.

\subsection{Replication Package\label{sec:reppack}}
All the data used in our study are publicly available.\footnote{See: http://doi.org/10.5281/zenodo.4589573} Specifically, we provide the R scripts that we used to extract the comments and perform the data analysis and processing, and the working data sets (phrases with TD type and user roles reporting them) used to run the statistical analysis.

\begin{figure*}[t!]
    \centering
    \includegraphics[width=\textwidth]{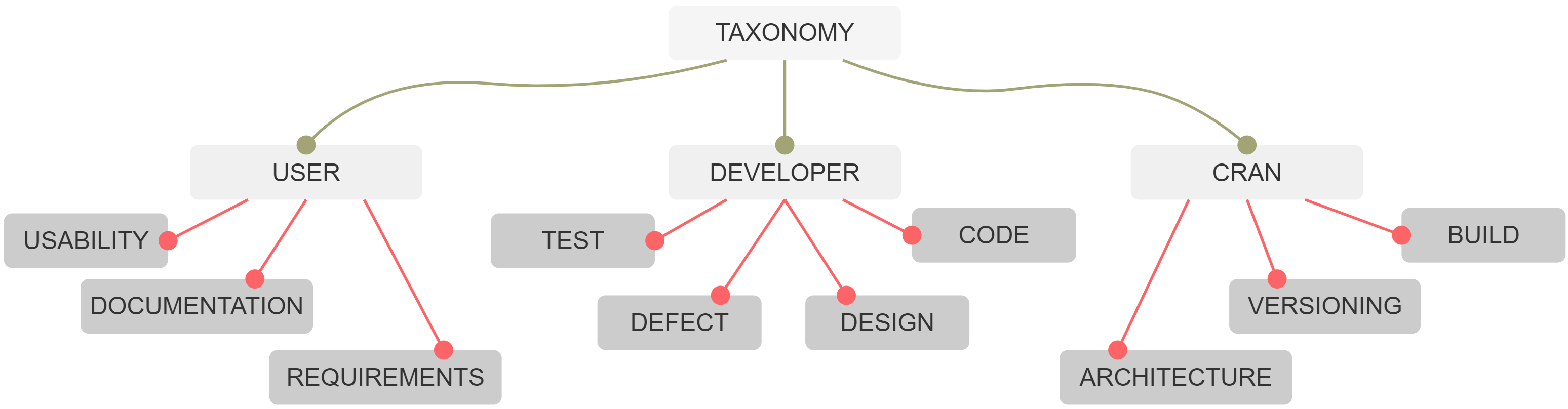}
    \caption{Summary of the proposed taxonomy of TD for R, grouping each \textit{type} (leaf level) in \textit{perspective} (mid level).}
    \label{fig:rq1_tree}
\end{figure*}

\section{Results \& Discussion \label{sec:results}}

The following subsections present the results organized by research questions and a brief discussion.

\subsection{RQ1: TD Taxonomy for R}
Table \ref{tab:rtd_taxonomy} presents the adapted definitions of the types of TD found through the manual analysis process (see Section \ref{sec:agreement}). The definitions are extracted from well-known academic sources, cited in the table, and then expanded or generalized to fit R concepts. For each type of debt, the table also presents an example extracted from the manual sample.

\begin{table*}[bp]
    \caption{Taxonomy of extended TD definitions for R, with an example extracted from the manually classified samples.}
    \label{tab:rtd_taxonomy}
    \begin{tabularx}{\textwidth}{l X X}
    \hline
    \textbf{Debt Type} & \textbf{R Definition} & \textbf{Example} \\
    \hline
    
    Architecture 
    &  Refers to the problems encountered in product architecture, for example, violation of modularity, which can affect architectural requirements (e.g. performance, robustness) \cite{brown2010managing,izurieta2012organizing}.
    & ``A common solution is to split the general tool and the specific application into separate packages.  This isn't required, but it is important to clearly separate the components of the code base and the documentation for both the EHR manipulation and the CCHIC use-case." \\
    
    \arrayrulecolor{lightgray}\hline
    
    Build 
    &  Refers to issues that make the build task harder, and unnecessarily time consuming. The build process can involve code that does not contribute to value to the customer. Moreover, if the build process needs to run ill-defined dependencies, the process becomes unnecessarily slow. When this occurs, one can identify build debt \cite{morgenthaler2012searching}. 
    
    In the context of R, build debt encompasses anything related to Travis, Codcov\@.io, GitHub Actions, CI, AppVeyor, CRAN, CMD checks, devtools::check.

    & ``No problems installing from a local tarball, but problems arose after installing using \texttt{devtools::install\_all()}"\\
    
    \hline
    
    Code 
    &  Refers to the problems found in the source code that can negatively affect the legibility of the code making it more difficult to maintain. Usually, this debt can be identified by examining the source code for issues related to bad coding practices \cite{bohnet2011monitoring}.
    
    In the context of R, code debt encompasses anything related to renaming classes and functions, $<$- vs. = , parameters and arguments in functions, FALSE/TRUE vs. F/T, print vs warning/message.

    & ``There were some instances where you used system variable names as local variable names in functions, e.g. line 36 in \texttt{auk\_time.r}, \texttt{time <- paste0(ifelse(nchar(time) == 4, "0", ""), time)}. It obviously doesn't cause an issue as is, but maybe it could down the line."\\
    
    \hline
    
    Defect 
    &   Refers to known defects, usually identified by testing activities or by the user and reported on bug tracking systems \cite{snipes2012defining}.
    
    & ``A similar strategy underlies \texttt{safe\_hex\_color()}, so the method has the same flaws. Plus, something mysterious happens sometimes:" \\
    
    \hline
    
    Design 
    & Refers to debt that can be discovered by analyzing the source code and identifying violations of the principles of good object-oriented design (e.g. very large or tightly coupled classes) \cite{izurieta2012organizing, seaman2011measuring}.
    
    In the context of R, design debt encompasses anything related to S3 classes and S4 methods, exporting functions with @export or the name pattern (visibility), internal functions with coupling issues, location of functions in the same file, selective importing @import (whole package) or @importFrom (a specific function), notations packageName::functionName and package:::function, returning objects (dataframes or tibbles), and Tidyverse vs. baseR.

    & ``**Internal functions** do not need to be exported (e.g. \texttt{parse.info}, \texttt{read.prov}). Remove the \texttt{@export} tag for these functions and add the tags \texttt{@keywords internal} and \texttt{@noRd}." \\
    
    \hline
    
    Documentation 
    &  Refers to the problems found in software project documentation and can be identified by looking for missing, inadequate, or incomplete documentation of any type \cite{Guo2011}.
    
    In the context of R, documentation debt encompasses anything related to Roxygen2 (e.g., @param @return, @example), Pkgdown, Readme files, and Vignettes.

    & ``It is not documented that the date-times are integers converted to POSIX times, so they are all after 1970." \\
    
    \hline
    
    Requirements 
    & Refers to trade offs made with respect to what requirements the development team needs to implement or how to implement them. Some examples of this type of debt are: requirements that are only partially implemented, requirements that are implemented but not for all cases, requirements that are implemented but in a way that does not fully satisfy all the non-functional requirements (e.g. security, performance) \cite{kruchten2012technical}.
    
    & ``Uploading multiples files+directories, however, is still very much a work in progress (i.e., not functional). It's something I would personally use quite a bit in my own work."\\
    
    \hline
    
    Test 
    &  Refers to issues found in testing activities that can affect the quality of those activities. Examples of this type of debt are planned tests that were not run, or known deficiencies in the test suite (e.g. low code coverage) \cite{Guo2011}.
    
     In the context of R, test debt encompasses anything related to coverage, covr, unit testing (e.g., testthat), and test automation.

    & ``Test coverage is good overall but sticks mostly to `happy path'. Untested error paths appear in \texttt{sync.R}, \texttt{utils.R}, \texttt{wget.R}, \texttt{postprocess.R} etc. I feel like a jerk for bringing this up. I know how unfun it is to write test cases for these." \\
    
    \hline
    
    Usability 
    & Refers to inappropriate usability decisions that will need to be adjusted later. Examples of this debt are lack of usability standard and inconsistency among navigational aspects of the software \cite{zazworka2013case}.
    
    In the context of R, test debt encompasses anything related to usability, interfaces, visualization and so on. 
    
    & ``A better error might say, \texttt{see ?ch\_generate} or print out the \texttt{all\_choices} vector." \\
    
    \hline
    
    Versioning 
    &  Refers to problems in source code versioning, such as unnecessary code forks \cite{greening2013release}.
    
    & ``I note only that four levels of versioning might be overkill - current version is \textit{0.0.4-1}. Up to the authors, but with just 3 levels you'd now be up to \textit{0.0.11} (or maybe \textit{0.2.x}), which still allows an approximately infinite future before version numbers get too high, and makes it easier for others to track version numbers."\\
    
   \arrayrulecolor{black}\hline
    \end{tabularx}
\end{table*}

The manual inspection highlighted three \textit{perspectives} that group the TD types in the domain of R programming in terms of `who' is affected by each type of debt the most. They were extracted during the manual inspection process considering R's intrinsic characteristics and previous research findings that linked different TD types together. They were devised according to `who' was affected by the debt, according to what the person wrote--this was obtained by exploring the context of each sentence. Often, they mentioned why something was problematic. For example, they would say, ``it is better for the user if...", or ``a user may find it confusing if...". Therefore, this classification into perspectives comes from the data, but there is a possibility that there are overlaps. Moreover, note that the reviews did not mention or reference TD explicitly. Figure \ref{fig:rq1_tree} presents the grouping as a graph, with perspectives in the mid-level and types of debts as leaves. These perspectives are:

\begin{itemize}
\item \textbf{User:} TD types grouped under this category are \textit{usability}, \textit{documentation} and \textit{requirements}. These debts ultimately affect mostly the user of the package -- i.e., the person who installs and imports the package into their own project -- than the developers. Previous work have established that \textit{documentation debt} is strongly related to (and sometimes caused by) \textit{requirement debt} \citep{Rios2020}. Moreover, other studies have demonstrated that poor implementation of requirements affects the \textit{usability} \cite{Lage2019}.

\item \textbf{Developer:} \textit{Design} and \textit{code debt} are often viewed as `two sides of the same coin' \cite{Buschmann2011}, intrinsically related in terms of code quality \cite{Zazworka2011}. A key point in common of \textit{code, design, test,} and \textit{defect} debt is that they are introduced and addressed during the development, testing and maintenance phases of the software development lifecycle \cite{snipes2012defining}. Previous studies have also demonstrated that developers focus on these four types of debt the most \cite{Codabux2013}, and that \textit{defect} tends to be introduced as a result of the other three types of debt \cite{Guo2011}.

\item \textbf{CRAN:} CRAN performs automated checks on package size, structure, and versioning, among others when a package is uploaded\footnote{\url{https://r-pkgs.org/r-cmd-check.html}}; thus, even if a package will not eventually be published there, it still goes through the automated check. Overall, these types of debt affect the maintainability of the software, which is what CRAN checks attempt to improve as well \cite{Claes2014}.\\
\end{itemize}

\begin{tcolorbox}
\textbf{Findings \#1.} We extracted ten different technical debt types from the peer-review documentation and proposed a taxonomy pertaining to three different perspectives, namely: user, developer, and CRAN.
\end{tcolorbox}

\subsection{RQ2: TD Distribution by Type}
\begin{figure}[h!]
    \centering
    \includegraphics[width=0.5\textwidth]{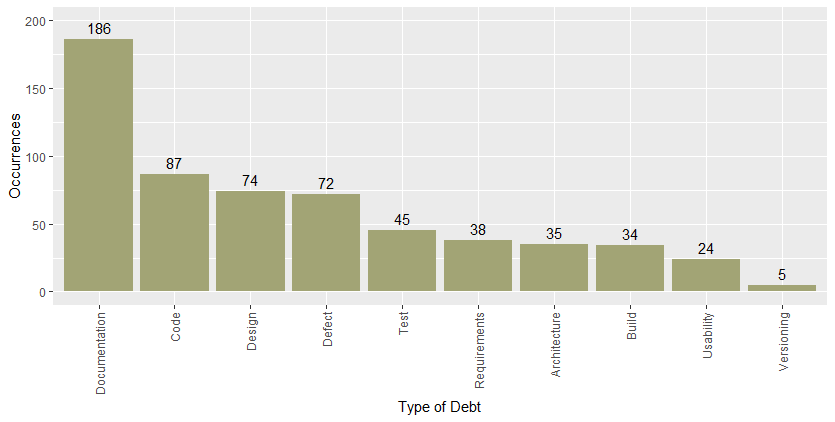}
    \caption{TD Distribution by Type}
    \label{fig:rq2}
\end{figure}

Figure~\ref{fig:rq2} depicts the distribution of the 600 phrases according to the ten TD types identified earlier. We found that the most prominent type of debt is \textit{documentation debt}. We had 186 occurrences out of 600, making up for almost 30\% of all the instances of debt. This is followed by \textit{code debt} (representing 15\% of the phrases) and \textit{design and defect debt} (each representing 12\% of the phrases). It is worth noticing that every occurrence of the same type of TD (if referring to different cases) is accounted for in this calculation.

 The other types of debt which were sparse were \textit{test debt, requirements debt, architecture debt, build debt, usability debt, and versioning debt}. They each represented 4$-$7\% of the phrases. \textit{Versioning debt} was the least reported type of TD  with only five instances out of 600. On average, each comment has six phrases which we encountered as TD, with a minimum of one, maximum of 50, and median of two. We did not notice any same TD phrase appearing in the same review. 
 
\begin{tcolorbox}
\textbf{Findings \#2.} Documentation debt is the most prominent type of debt, while versioning debt is the least prominent type of debt in the dataset.
\end{tcolorbox}

\subsection{RQ3: Association of User Roles and TD Types}
We manually reviewed the issues' comments to identify the user roles. This manual investigation ensures that we extract all the roles for a comment since a user can have different roles in different issues. Next, we use the comment's unique identifier to label the roles for each of the 600 phrases. The distribution of phrases for user role is shown in Table \ref{tab:TD_roles}. 

\begin{table}[h!]
  \caption{Technical debt counts by user roles}
  \label{tab:TD_roles}
  \centering
  \begin{tabularx}{0.4\textwidth}{Xc}
\hline
Role  & TD Phrase Count \\
\hline
\textbf{op\_author} & 48\\

\textbf{handling\_editor} & 68 \\

\textbf{author} & 5\\

\textbf{editors} & 5\\

\textbf{reviewers} & 474\\

\hline    
    \end{tabularx}
\end{table}

As the number of phrases written by \textit{authors} and \textit{editors} are low, we combined the \textit{author} with \textit{op\_author}, and the \textit{editors} with \textit{handling\_editor} phrase counts. The \textit{op\_author} and \textit{author} are developers of a package; and the responsibilities of \textit{handling\_editors} and \textit{editors} are also similar to each other. For RQ3, we refer to these combined roles as \textit{author} and \textit{editor} respectively. This combination is also confirmed by one of the current editors of the \textit{rOpenSci}. Therefore, combining these roles will not affect the reliability of the results.   

We applied the Chi-Square test and Cramer's V to find the association between the three user roles (i.e. \textit{author}, \textit{editor}, and \textit{reviewer}) and the TD  types. The Chi-Square test is used to find the independence among two categories, where each category has multiple levels (more than two), and Cramer's V shows the strength of this association \cite{mcdonald2009handbook}.  
A main assumption of the Chi-Square test is that the expected frequencies in the contingency table are not less than 5 for more than 20\% of the cell values \cite{mcdonald2009handbook, kroonenberg2018tale}. This assumption is not satisfied in our case, as the number of phrases labelled as \textit{design}, \textit{versioning}, and \textit{usability} by \textit{authors} and \textit{editors} are low. Therefore, we use the three TD perspectives discussed earlier (i.e. User, Developer, and CRAN) instead of the TD types. 
Our null hypothesis is as follows:

\textit{Null Hypothesis $H_0$}: The TD types in the comments do not depend on the user role associated with the comment's writer.

This hypothesis is rejected with a $p$-value of $4.2E-08$ for $\alpha = 0.05$ and a Cramer's V value of $0.18$.
Therefore, the TD types and the user roles are dependent but with a small association. 
We also applied Post Hoc tests \cite{mcdonald2009handbook} to identify the differences between the user roles pairwise: \textit{author} vs. \textit{editor}, \textit{author} vs. \textit{reviewer}, and \textit{editor} vs. \textit{reviewer}. As the number of comparisons is more than two, Bonferroni Correction \cite{mcdonald2009handbook} is applied to adjust the alpha value ($\alpha = 0.016$). 
The test results show that there is not a significant difference between the \textit{author} and \textit{editor} in terms of TD types in the comments they wrote. However, the \textit{reviewer}'s role affect the TD types we identified in the comments. A significant difference is found between the \textit{editor} and \textit{reviewer} roles, and \textit{author} and \textit{reviewer} roles with $p$-values of $5.5E-04$ and $5.8E-08$, respectively. The effect size (i.e. strength of the association) using Cramer's V score shows a small association between the \textit{editors} vs. \textit{reviewers} ($0.16$) and a medium association among the \textit{authors} and \textit{reviewers} ($0.25$). 

\begin{figure}[h!]
    \centering
    \includegraphics[width=0.5\textwidth]{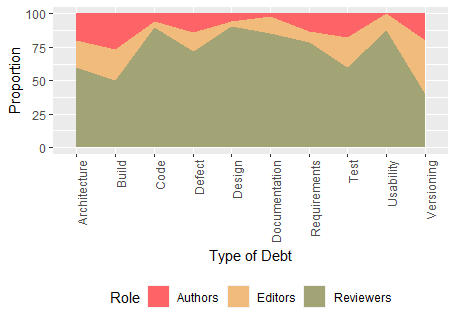}
    \caption{The Percentage of TD Types Compared by User Roles}
    \label{fig:rq3_A}
\end{figure}

Figure \ref{fig:rq3_A} represents the percentage of TD phrases in each category, where we separated all phrases of each TD by user roles. For all TD types except \textit{versioning debt}, \textit{reviewers} have the highest number of comments, with \textit{design debt}, \textit{code debt}, and \textit{usability debt} having the highest percentages, 90.5\%, 89.6\%, and 87.5\% respectively. This is followed by 85.4\% for \textit{documentation debt}, 79\% \textit{requirements debt}, and 72\% for \textit{defect debt}. \textit{Test debt}, \textit{architecture debt} and \textit{build debt} reported by \textit{reviewers} are 60\%, 60\%, and 50\% respectively. 
40\% of the \textit{versioning debt} that we identified are reported by the \textit{reviewers}, 40\% by the \textit{editors} and 20\% by the \textit{authors}. 

The percentage of the phrases identified as \textit{documentation debt}, \textit{test debt}, and \textit{usability debt} written by \textit{editors} are the highest (12\%, 22\%, and 12.5\% respectively). The next highest are the ones written by the \textit{reviewers}.
We found a small percentage of \textit{documentation debt} and no instances of \textit{usability debt} among the phrases written by \textit{authors}. 
The percentage of phrases with \textit{requirements debt} written by the \textit{authors} is approximately twice as many as the \textit{editors}. 
The percentage of phrases classified as \textit{architecture debt} and \textit{defect debt} are the same for \textit{authors} and \textit{editors} - 20\% and 13.8\% respectively. The percentages for the \textit{authors} and \textit{editors} are very close for \textit{design debt} and \textit{code debt} (4-6\%). The highest percentages belong to \textit{build debt} with 23\% and 26\% encountered by \textit{editors} and \textit{authors} respectively.

\begin{figure}[h!]
    \centering
    \includegraphics[width=0.5\textwidth]{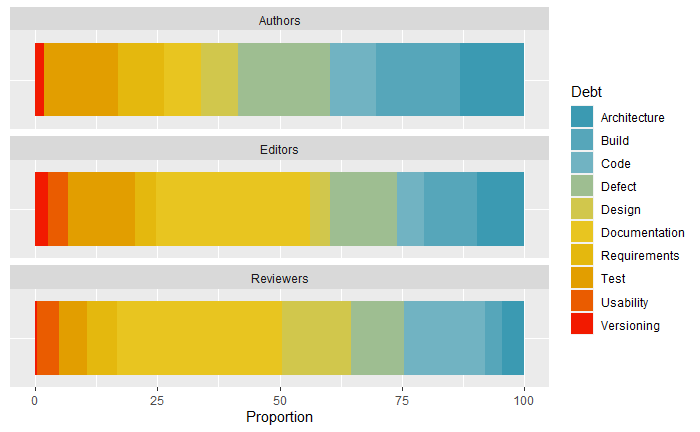}
    \caption{The Distribution of TD Types Found in the Phrases Written by Each User Role (Shown in Percentages)}
    \label{fig:rq3_B}
\end{figure}

We also separated the phrases written by each user role to determine the percentage of each TD type for each user role. These TD distributions are shown in Figure \ref{fig:rq3_B}. The \textit{authors} have \textit{defect debt} as the highest TD (18.8\%) instances reported, followed by \textit{build debt}, \textit{test debt}, and \textit{architecture debt} with 16.9\%, 15\%, and 13.2\%, respectively. The next two highest percentage of the TD phrases for \textit{authors} belong to \textit{code debt} and \textit{requirements debt} with 9.4\% each. There is no phrases with  \textit{usability debt} and the \textit{versioning debt} percentage is also very small ($<$ 2\%) for the \textit{authors}. The debt instances identified by the \textit{editors} are mostly \textit{documentation} (31.5\%), followed by \textit{defect} and \textit{test} (13.7\% each), \textit{build} (10.9\%) and \textit{architecture} (9.5\%). The percentages of phrases with \textit{code}, \textit{design}, \textit{requirements}, and \textit{usability} are almost the same (4-5 percent). \textit{Versioning} is the least reported (2.7\%) by \textit{editors}. \textit{Reviewers} report mostly \textit{documentation} (33.5\%), followed by \textit{code}, \textit{design}, and \textit{defect} with 16.4\%, 14.1\%, and 10.9\% respectively. The percentage for the other categories is less than 10\%. Similar to other roles, \textit{versioning} is the least reported with only 2 instances (approx. 0.4\%). 

\begin{tcolorbox}
\textbf{Findings \#3.} The user roles affect the quantity and the TD types that are reported. The \textit{reviewers} report more TDs than other user roles. Moreover, the type of TD encountered by each user role is different. The \textit{reviewers} focus on the \textit{documentation}, \textit{code} and \textit{design} debts more than others. The \textit{authors} concentrated mostly on \textit{defect}, \textit{test}, and \textit{architecture} debts. The \textit{editors} pay more attention to \textit{documentation}, \textit{defect}, and \textit{requirements} debts. 
\end{tcolorbox}

\subsection{Discussion}
Our findings show the number of instances for each TD type encountered by each user role differ greatly. For example, in our sampled data, the \textit{authors} pay more attention to \textit{defect debt} and less attention to \textit{usability} and \textit{versioning debts}. 
This can be due to the familiarity of the developers with their own package and being aware of potential defects and limitations. 
The \textit{editors} and \textit{reviewers} on the other hand, pay more attention to the \textit{documentation debt} (above 30\% of their phrases is identified as documentation). This aligns with the reasons for the creation of R. R is a community created with the intention of being fully open source \cite{Gentleman2004, Ihaka2017}. As a result, its focus on documentation is well-known \cite{Wickhan2017}, as well as the emphasis on the maintainability of contributed packages -- for example, CRAN automatically compiles the package's documentation to provide a PDF-style document alongside the package \cite{Claes2014}. 
The results mentioned above are an eye-opener for each of the user roles on \textit{rOpenSci}. Although managing technical debt is vital to the software quality, not all users give the same importance to the different types of debt. As we expected, the \textit{reviewers} reported more TDs compared to others. We relate this result to the fact that the reviewers' responsibility is to ensure that the approved packages achieve a certain quality. This is reflected both in the high number of TD phrases in our study that are written by \textit{reviewers} and the significant difference of the distribution of TD types of \textit{reviewers} and the two other roles. 

The taxonomy of TD proposed in this paper is based on existing proposals for OO Programming. As a result, the definitions had to be adjusted to fit the characteristics of R, a functional dynamically-typed language. However, it is worth noting that several types of TD were not found in the discussions ---examples are \textit{service} and \textit{installation debt}. This may be due to either difference in the language or the ``topic scope'' of \textit{rOpenSci} which does not provide the opportunity to report such types of debt.

Other types of debt, specific to scientific software have been proposed in the literature, such as \textit{algorithm} \cite{liu2020usingSATDinDL} or \textit{reproducibility debt} \cite{sculley2015hiddenTDinML}. However, these definitions have been lightly explored as they are recent proposals. A more in-depth study of R packages should be conducted to understand their impact on R fully. It is also worth noting that the proposed taxonomy and the frequencies of the different debt types may be different from those reported in SATD studies. SATD occurs when developers admit poor solutions in the source code comments \cite{SIERRA201970}. Possible differences between the debt types in our work and the SATD studies may be hypothesized to be centered around the presence of other types of debt such as \textit{code} and \textit{design debt}.

\section{Implications \label{sec:implications}}
\textbf{Implications for Researchers.} 
Our study shows that not all types of TD  are equally prevalent in R packages. Our findings reveal that \textit{documentation debt} is the most recurrent type of debt. In general, we know that documentation debt is one of the least studied types of debt. This reinforces the need for \textit{documentation debt} to be studied more extensively. For instance, there is a need to establish what constitutes good documentation to minimize \textit{documentation debt} or what guidelines to follow when writing software documentation, both traditional and scientific. This is the first study to explore a taxonomy for TD  for scientific software to the best of our knowledge. We conducted our study in the context of R. This provides opportunities for researchers to investigate TD  in other scientific software using our taxonomy as a starting point and augment it. The fact that comments from the different user roles focus on different types of debt is intriguing. \textit{Reviewers} seem more concerned with creating a user-friendly package in terms of documentation, while \textit{authors} seem focused on the package's structure. Further studies need to be conducted to understand the rationale behind these distinctions.

\textbf{Implications for the R Community.} 
To the authors' knowledge, this is the first taxonomy for R packages, including tailoring existing definitions to the intrinsic characteristics of the R language. For developers, this will allow the specification and elaboration of specific smells in the future, leading to better coding practices that will enhance the quality of the available packages. Moreover, these concepts can also be used when training new developers (i.e., by organizing workshops). Regarding the peer-review of R packages, the revision process (and specifically, the templates used by the referees) can be upgraded to include points that tackle those types of TD that are less encountered. The submission template used by authors can also be altered to emphasize on documentation and replication, as it is essential for the usability of the packages. Moreover, the results of our study reveal the importance of the reviewers' role in reviewing packages. Although none of the users discuss TD directly, nor mention the metaphor, their commitment to quality is perceived through their comments.  Although we did not directly trace the application of this feedback in the source code, this is validated through the review process of \textit{rOpenSci}. Based on the number of instances found for each TD category, we recommend that researchers and practitioners require new solutions (such as updating the peer-review criteria and updating CRAN checks) to ensure all TD types are inspected in the peer-review process of R packages. Currently, the most prevalent debt types encountered are \textit{documentation} and \textit{requirements}, and the ones with least occurrences are \textit{usability} and \textit{versioning}. So, we suggest researching new ways to check how \textit{usability} and \textit{versioning} technical debts are addressed in R package peer-reviews.

\section{Threats to Validity \label{sec:threats}}
\textbf{External Validity.} External validity refers to the ability to generalize results. Our study was conducted on the documentation of the peer-review of R packages. We cannot draw general conclusions about the TD  in other scientific software. Despite having a small sample size to start with (95 out of the 358 comments had TD), the comments were quite lengthy and needed to be broken down into phrases, each one pertaining to a different type of debt. Instead, we ended up with a sample of 600 phrases or instances of TD. However, the goal of our study is not to provide an analysis of TD  that applies to all scientific software but to serve as a baseline for future studies in scientific software analysis or peer-reviews. It is worth noting that the results are based on the whole ecosystem of rOpenSci, but similar peer-review processes have been used in other communities (i.e. BioConductor, pyOpenSci and Journal of Open Source Software). Therefore, our results can be applied to those communities. However, these results would not be applicable to OO or functional languages.

\textbf{Internal Validity.} This threat refers to the possibility of having unwanted or unanticipated relationships. As we selected a sample of review comments in our analysis, sampling bias might impact our conclusions. To mitigate this threat, we randomly selected the comments with a confidence level of 95\% and a confidence interval of 5 for our classification and did not have any restrictions on the timeline for the extraction. Therefore, our results are not bound to a certain period. In addition, to build our taxonomy, we used types of TD  previously defined in the literature as a starting point. However, these debt types have been used extensively in both research and practice and validated by many studies. 

Moreover, the manual process of classifying the comments according to debt types can be biased. To mitigate this threat, the classification was performed by two of the authors separately and then compared to finalize the classification. The card sorting method might be a potential source of bias, and despite the rigor followed by the authors to classify the comments, replication of this study might lead to different results. To mitigate this threat, the authors thoroughly discussed the conflicts and assigned the final debt type to the phrases which were the source of conflicts. Lastly, as we are using a different artifact in the format of the documentation of the peer-review of R packages as our dataset, there might be a threat associated with the authors being unfamiliar with the dataset as they usually deal with artifacts from traditional software development such as source code or code comments. To alleviate potential problems that this might introduce in our analysis, we used 100 records of the comments as a learning phase. This learning phase ensures that our authors become familiar with the dataset and the discussions for R package peer-reviews and ensure that the authors agree on the context of TD in each category.  

\textbf{Construct Validity.} Construct validity refers to the degree to which a test measures what it claims to be measuring. Some of the debt types had a low number of occurrences after the classification and can be a potential bias for our findings in RQ3. However, the Chi-Square test does not have any assumption about the normality or the frequency of the sample data, and therefore, the test results are reliable. This ensures the association between the user roles and TD types, with an effect size (i.e. Cramer's V score). Besides, to further analyze the results and find out the differences among each pair of the TD type distributions, we applied Chi-Square Post Hoc Tests. This step identifies the pair-wise comparison among the distributions for user roles, reported in the results section. 

\section{Conclusions and Future Works \label{sec:conclusions}}
In this study, we manually analyzed the technical debts in the documentation of R packages' peer-review comments. We proposed a taxonomy of ten different TD types pertaining to three different perspectives. Our findings reveal that some technical debts are prominent, and some appear with a low frequency. These TD occurrences significantly differ according to different user roles. The results of the statistical tests also confirm the dependency and differences between the TD types and user roles. The proposed taxonomy, investigating TD in the comments of the peer-review R packages, and differences among user roles are the main contributions of this study. In addition, we provide access to our scripts and datasets in the replication package (see Section \ref{sec:reppack}). We conclude with some implications for researchers and practitioners in the R community based on our findings. 

The next step is to replicate this study but in a different domain. We plan to analyze the technical debt in \textit{BioConductor} packages to investigate the differences among domain-specific and more general package reviews and extend our findings to be as generalizable as possible. \textit{BioConductor} only onboards packages oriented to work with genomics data, while  \textit{rOpenSci} has a wider range of on-topic packages, including data munging, API connections, and geospatial-related packages. Following that, we will automate the identification of TD types in the comments of peer-review packages in both R and the \textit{BioConductor} domain. 



\section{Acknowledgements}
We thank Scott Chamberlain, Noam Ross and Karthik Ram, Associate Editors of \textit{rOpenSci} and the MSR reviewers.

\bibliographystyle{IEEEtranN}
\bibliography{references}

\end{document}